\documentclass[aps,prl,showpacs]{revtex4}

\begin{document}

\title
{Casimir effect for the sphere revisited}

\author {C. R. Hagen\cite{Hagen}}

\affiliation {Department of Physics and Astronomy\\ University of
Rochester\\ Rochester, N.Y. 14627}

\begin{abstract}
In a recent work Brevik \emph{et al.} have offered formal proofs
of two results which figure prominently in calculations of the
Casimir pressure on a sphere.  It is shown by means of simple
counterexamples that each of those proofs is necessarily
incorrect.
\end{abstract}

\pacs {42.50.Lc, 03.50.De, 12.20.-m, 03.70.+h} \maketitle

\begin{center}
        {\bf I INTRODUCTION}
        \end{center}
\  \\

The Casimir energy of a conducting sphere has been considered in a
number of papers in the past several years.  In particular this
author [1] reexamined the well known calculation of Milton,
DeRaad, and Schwinger [2] and raised certain concerns about the
means used in that work to carry out those calculations.  These
may be conveniently referred to as the pressure-stress relation
and the claimed validity of a certain contour rotation.

Subsequent to the publication of ref.1, Brevik, \emph{et al.} [3]
have stated their strong objections to the conclusions reached in
that work and have offered what they claim to be proofs of two
results which support the conclusions of ref. 2.  As is shown in
the present work neither of those proofs can be correct since
counterexamples can be shown to exist.  In {\bf II} the
pressure-stress issue is discussed, it being shown that a proof
based merely on the assumptions of ref. 3 is logically impossible.
Section {\bf III} similarly provides a simple counterexample to
the contour rotation result of refs. 2 and 3.

Before proceeding to these two main results it may be appropriate
to comment briefly on a third issue discussed in ref. 1 having to
do with the issue of boundary conditions.  As is well known any
calculation based on quantum field theory must utilize so-called
causal boundary conditions (i.e., positive frequencies in the
future and negative frequencies in the past).  Reference 2,
however, is based (by its own claim) on outgoing wave boundary
conditions, which according to standard usage would imply
functional dependence of the form $\omega (r-t)$.  This is quite
different from the result identified as outgoing wave boundary
conditions as given by Eq.(1) of ref. 3.  It is therefore clear
that this issue has its origin in imprecise language rather than
any significant issue of physics.
\\

\begin{center}
        {\bf II THE PRESSURE-STRESS CONNECTION}
      \  \\
      \end{center}

In ref. 2 and a number of subsequent related applications the
Casimir pressure on a sphere is computed by evaluating the
discontinuity of the $T^{rr}$ component of the stress tensor, an
approach which has a considerable amount of intuitive appeal.
What is not clear, however, is whether this calculational method
has a rigorous basis.  What is shown here (the claim of ref. 3
notwithstanding) is that one cannot hope to establish the
legitimacy of this approach making recourse only to the
appropriate divergence relation for the stress tensor.

The proof of ref. 3 argues from the result
\begin{equation}
\nabla_\alpha T^{\alpha\beta}=-F^{\beta\lambda}J_\lambda =
-f^\beta
\end{equation}
where $f^\beta$ is identified as the four-vector force density and
is nonzero only within the conducting shell [4].  It is then
inferred that this force is necessarily identical to the Casimir
force, thereby leading to the conclusion that the Casimir pressure
is indeed given by the discontinuity of $T^{rr}$.  It is easy to
see that this is nothing more than proof by definition--namely,
that because $f^{\beta}$ has certain properties characteristic of
a force, it \emph{must} be the Casimir force.

To display the contradiction, one makes reference to the
corresponding stress tensor for a spin-zero field $\phi$ which has
the form
\begin{equation}
T^{\mu\nu}=T^{\mu\nu}_1 + T^{\mu\nu}_2
\end{equation}
where

$$T^{\mu\nu}_1= \phi^\mu\phi^\nu-{1\over
2}g^{\mu\nu}\phi_\alpha^2$$

$$T^{\mu\nu}_2={1\over
6}[g^{\mu\nu}\partial^2-\partial^\mu\partial^\nu]\phi^2$$ and
$\phi_\mu=-\partial_\mu\phi$.  Note that $T^{\mu\nu}$ is the
so-called ``improved" (i.e., traceless) stress tensor, while
$T^{\mu\nu}_1$ is the canonical (i.e.,``unimproved") version.  One
can now include a coupling to a scalar current $J$ (the analogue
of $J^{\mu}$ of Eq.(1)) to obtain
\begin{equation}
\nabla_\alpha T^{\alpha\beta}=-\phi^\beta J=-f^\beta
\end{equation}
which therefore (in accord with the approach of ref. 3) must imply
that the rhs of (3) is the Casimir force for the scalar case.
However, it is important to note that the same relation holds for
the unimproved tensor $T^{\alpha\beta}_1$ as well, and it
consequently follows that the Casimir pressure can be obtained
from the discontinuity of either $T^{\alpha\beta}$ or
$T^{\alpha\beta}_1$.  Unless these have the same discontinuity,
one must conclude that the result fails in (at least) one of the
two cases in which the proof of ref. 3 is extended to the scalar
case.

To demonstrate such a failure one makes recourse to the case of
the spherical shell $a<r<R$ and seeks solutions which satisfy the
boundary condition $\partial_rr\phi=0$ at the boundaries $r=a, R$.
The issue to be resolved is whether the Casimir pressure
$$p=-{1\over 4\pi a^2}{\partial E\over\partial a}$$ where $E$ is
the Casimir energy $$E=\int d{\bf x}T^{00}({\bf x})$$ is given by
the negative of $T^{rr}(r=a)$ for each allowed mode of the system
[5].  Since the aim here is to provide a demonstration using the
simplest possible choice of $\phi$, it is convenient to take the
spherically symmetric free field solution
$$\phi=A{\cos[\omega_n(r-a)]\over \omega_nr}.$$ The eigenmodes
$\omega_n$ are readily seen from the boundary conditions to be
given by $\omega_n={n\pi\over R-a}$ while $A$ is determined from
the condition

$$2\omega_n A^2\int d{\bf x}\phi^2=1.$$

Using $T^{\mu\nu}_1$ and $T^{\mu\nu}_2$ successively one finds
without difficulty that the two ``Casimir energies" $E_1$ and
$E_2$ are given by $${1\over 4\pi}E_1={1\over 2}\omega_n+{1\over
2n\pi}({1\over a}-{1\over R})$$ and $${1\over 4\pi}E_2=-{1\over
3n\pi}({1\over a}-{1\over R}).$$ One has the corresponding results
for $T^{rr}(r=a)$ $$T^{rr}_1(r=a)={n\pi\over 2a^2(R-a)^2}+{1\over
2n\pi a^4}$$ and $$T^{rr}_2(r=a)=-{2\over 3n\pi a^4}.$$ Using
these results it is straightforward to show that for both
$T^{\mu\nu}_1$ and $T^{\mu\nu}_2$ the pressure-stress relation
fails, but that it does hold for the sum (i.e., for the
``improved" case).

On the basis of these results one can conclude that the proof of
the pressure-stress relation claimed in [3] is not in fact valid.
In general it simply cannot be possible to prove such a relation
solely on the basis of arguments deriving from the divergence of
the stress tensor.  It may, of course, be possible to do so by
including as well the requirement that the stress tensor be
traceless.  However, that property was not invoked in [3] and thus
the result quoted there is necessarily incorrect.

For purposes of clarity it is certainly appropriate to remark here
that the claimed proof of ref.3 deals only with the case of the
electromagnetic field.  Yet, having said that, it is striking that
(as shown here) virtually identical techniques imply a
contradiction when applied to the scalar field case.  In addition
it  should not be overlooked that section {\bf III} of ref.3
presents a scalar field Casimir calculation which implicitly
assumes the pressure-stress relation despite the demonstrated
failure of the proof in that case.
\\

\begin{center}
        {\bf III CONTOUR ROTATION}
        \end{center}

\  \\

A significant problem encountered in ref. [2] (and in subsequent
related works) is the evaluation of integrals with rapidly
oscillating integrands.  It is argued there that this can
generally be accomplished by a contour rotation which results in a
much more manageable integrand.  The mathematical basis for that
rotation is given in [2] and repeated virtually verbatim in [3].
It is shown here by an explicit calculation that the mathematical
steps in that rotation must necessarily be incorrect.

This is easily accomplished by considering the $l=0$ contribution
[5] to the integral in Eq.(2) of ref. [3].  To within an
uninteresting normalization factor one can write that part of
Eq.(2) as $$f={ia\over 2}\int_C d\omega e^{-i\omega \tau} \left(
ka \left[ { H^{(1)\prime}_{1/2}(ka)\over H^{(1)}_{1/2}(ka) } + {
J^\prime_{1/2}(ka)\over J_{1/2}(ka)} \right]+1\right)$$ where
$k=|\omega|$ and $C$ is a contour just above the real axis for
$\omega>0$ and just below the real axis for $\omega<0$.  It is
claimed in [3] that as a result of contour rotation this can be
transformed into $$f_E = {-1\over 2}\int_{-\infty}^{\infty} dy
e^{i\delta y} \left( x{K^\prime_{1/2}(x) \over
K_{1/2}}+x{I^\prime_{1/2}(x)\over I_{1/2}(x)}+1 \right)$$ where
$x=|y|$.  The content of this claim lies in the fact that, if
true, $f$ and $f_E$ must be equal in the limit of vanishing cutoff
(i.e., in the limit $\tau,\delta\to 0$).  It is the advantage of
the $l=0$ case that both of the above integrals can be evaluated
analytically.  The latter is particularly simple, with the result
being $$f_E=-{\pi^2\over 12}.$$

To evaluate $f$ one notes that the Bessel functions of order
${1\over 2}$ allow it to be written as $$f={i\over 2}\int_C dx
\exp[i|x|-ix{\tau\over a})] {x\over \sin x}.$$ It is convenient to
break the integral into a principal value term and a sum over
contributions from the poles of the integrand.  Upon taking the
limits of the integral as $\pm R$ where $R=(N+\xi)\pi$ with $N$ a
large integer and $0<\xi<1$, the real part of this integral can
readily be evaluated.  For small $\tau$ the result is $$\Re
f=-{\pi^2\over 12}+{\pi^2\over 2}\cos (N\pi\tau/a)
[N(1-2\xi)-\xi^2+{1\over 6}].$$ Since this has no well defined
$N\to \infty$ limit, one concludes by this direct calculation that
the original integral $f$ does not agree with the result for
$f_E$.  It is not difficult to show that this occurs precisely
because the integrals along the quarter circles of radius $R$ in
the first and third quadrants fail to vanish in the $R\to \infty$
limit.   At least for the $l=0$ case it appears that the use of
the alternative cutoff $e^{-k|x|}$ with $k>0$ could formally avoid
this difficult issue of contour rotation.  However, such a cutoff
would  require significant modification of the approach of refs.
[2] and  [3] where the cutoff originates in the underlying
Minkowski space formulation of the theory.
\\

\begin{center}
        {\bf IV CONCLUSION}
\end{center}
\   \\

In this work two alleged theorems of ref. 3 have been shown by
specific counterexamples to be incorrect.  The first of these has
to do with the pressure-stress relation, and it has been
demonstrated here that the proof offered in [3] \emph{must} fail
since the spin-zero canonical stress tensor is in specific
disagreement with the claimed pressure-stress relation.  If such a
relation can in fact be established, its proof necessarily must
include the property of tracelessness, an aspect which nowhere
appears in the proof of ref. 3.  Similarly, the contour rotation
result supposedly proved in [2] and then again in [3] has been
shown to fail for the case of $l=0$, the only instance for which
it appears possible to do an exact calculation.

A final comment has to do with the rather extensive set of remarks
made in ref. 3 concerning the experimental side of the Casimir
effect.  These were apparently intended to rebut the two sentences
of ref. 1 (described in [3] as \lq\lq objectionable") which
remarked that some recent experiments [6] might be viewed as less
than compelling evidence.  This had to do with the fact that
strict plane parallel plate geometry has not been utilized in
those experiments, and the correction to a spherical lens has not
been rigorously carried out. It remains true that those
corrections have not been calculated beyond a reasonable doubt.

\acknowledgments

This work is supported in part by the U.S. Department of Energy
Grant No.DE-FG02-91ER40685.


\begin{thebibliography}{99}
\bibitem{notes} It is appropriate to remark here that the statement
in ref. 3 to the effect that ref. 1 assumes that $\nabla_\alpha
T^{\alpha\beta}=0$ everywhere is simply incorrect. Such an
assertion would be equivalent to the insupportable claim that the
stress tensor has \emph{no} discontinuity at the surface of the
sphere.
\end{thebibliography}
\end{document}